\documentclass[12pt]{iopart}

\usepackage{graphicx}
\DeclareGraphicsExtensions{.eps,.ps}
\begin{document}

\def\bbm[#1]{\mbox{\boldmath$#1$}}

\title{Frictional quantum decoherence}

\author{Bruno Bellomo\dag\, Stephen M. Barnett\ddag\  and John Jeffers\ddag}

\address{\dag\ Dipartimento di Scienze Fisiche ed
Astronomiche dell'Universit\`{a} di Palermo, Via Archirafi, 36,
I-90123 Palermo, Italy.}

\address{\ddag\ Department of Physics, University of Strathclyde,
Glasgow G4 0NG, United Kingdom}

\begin{abstract}
The dynamics associated with a measurement-based master equation
for quantum Brownian motion are investigated. A scheme for
obtaining time evolution from general initial conditions is
derived. This is applied to analyze dissipation and decoherence in
the evolution of both a Gaussian and a Schr\"{o}dinger cat initial
state. Dependence on the diffusive terms present in the master
equation is discussed with reference to both the coordinate and
momentum representations.
\end{abstract}

\submitto{\JPA} 

\section{Introduction}

The connection between quantum dissipation and decoherence is a
topic of longstanding interest \cite{Caldeira-Leggett 1983,Hu
1992,Sandulescu 1987,Gallis 1993,Joos 1985, Diosi 1995,Horneberger
2003,Scott 2001,Mensky 2003}. The main systems analyzed in this
perspective are the damped harmonic oscillator, two level systems
and quantum Brownian motion. For such systems the Hamiltonian
description is not appropriate, and the most successful results
come from the reduced description of a particle interacting with
some type of reservoir \cite{Petruccione-Breuer libro 2002}.
Classical understanding of the phenomenon is well-established,
relying on Langevin or Fokker-Planck equations obtained by
considering a particle interacting with a bath of independent
oscillators \cite{Zwanzig 1973}. The quantum counterpart of
classical Brownian motion, however, has only recently been cast
into standard equations \cite{Barnett 2005,Barnett 2006}.

Several approaches have been followed in order to obtain a quantum
description of the dynamics of the Brownian particle:
\begin{itemize}
    \item a model-reservoir approach, leading to the famous Caldeira and
Leggett master equation, which assumes the particle to be coupled
to an environment described by a collection of simple harmonic
oscillators \cite{Caldeira-Leggett 1983,Hu 1992}, and into which
suitable terms can be added in order to produce a satisfactory
Markovian equation of the required Lindblad form \cite{Sandulescu
1987,Gallis 1993},
    \item a  dynamical approach modelling more closely the random
collisions between the Brownian particle and the particles that
make up its surrounding environment \cite{Joos 1985,Diosi
1995,Horneberger 2003},
    \item a measurement-based approach focusing on the information about
the Brownian
    particle carried away by the particles of the medium during the
collisions. This information is available, at least in principle,
by monitoring the environment particles \cite{Scott 2001,Mensky
2003,Barnett 2005,Barnett 2006}.
\end{itemize}

The property of complete positivity, to be satisfied by a master
equation for the reduced density operator of the particle, is a
useful and stringent requirement in the study of subdynamics in
quantum mechanics \cite{Lindblad 1976}. The various approaches are
described in \cite{Barnett 2005}, where the results obtained with
the different methods are discussed. Of particular relevance is
whether or not the proposed master equations are Markovian and of
Lindblad form.

Here the approach that we use to describe the quantum Brownian
particle dynamics is the measurement-based one found in
\cite{Barnett 2005,Barnett 2006,Cresser 2006}. The collisions with
the surrounding particles are considered to perform a random
sequence of measurements feeding information about the position
and momentum of the Brownian particle into the environment. Even
if both of these quantities cannot be known with total precision
at the same time, it is possible to simultaneously measure
position and momentum by introducing some degree of imprecision
for both. Using non-quantum-limited measurement techniques to
represent the acquisition of this information \cite{Busch 1995}
has led to the following master equation, in the limit of frequent
collisions which make very weak joint measurements of position and
momentum \cite{Barnett 2005}:
\begin{equation}\label{starting master equation}
\frac{\mathrm{d} \hat{\rho}}{\mathrm{d}
t}=-\frac{i}{\hbar}\left[\frac{{\hat{p}}^2}{2M},\hat{\rho}
\right]-\frac{i \gamma}{2
\hbar}\left[\hat{x},\{\hat{p},\hat{\rho}\} \right]-\frac{D_{pp}}{
\hbar^{2}}\left[\hat{x},[\hat{x},\hat{\rho}] \right]
-\frac{D_{xx}}{ \hbar^{2}}\left[\hat{p},[\hat{p},\hat{\rho}]
\right]\, .
\end{equation}
This equation is of the required Lindblad form provided that
\begin{equation}\label{Lindblad condition}
D_{pp}D_{xx}\geq (\hbar \gamma /4)^2 \,.
\end{equation}
Here $M$ is the mass of the Brownian particle, $\gamma$ is the
damping coefficient while $D_{pp}$ and $D_{xx}$ are diffusion
coefficients given by
\begin{equation}\label{dipp}
D_{pp}=\gamma \left[Mk_BT+\frac{m}{M}{(\Delta_\sigma p)}^2\right]+
\frac{R \hbar^{2}}{8 {(\Delta_\sigma x)}^2}\, ,
\end{equation}
and
\begin{equation}\label{dixx}
D_{xx}=\frac{ R \hbar^{2}}{8 {(\Delta_\sigma p)}^2}\, .
\end{equation}
The particles forming the environment have mass $m$, $R$ is the
average rate of collisions and $\Delta_\sigma p$ and
$\Delta_\sigma x$ represent the increase in the standard
deviations due to the measurements of position and momentum over
and above the intrinsic variances \cite{Barnett 2005}.

Satisfaction of the condition of Eq.\,(\ref{Lindblad condition})
using Eq.\,(\ref{dipp}) and (\ref{dixx}) ensures that the master
equation is of Lindblad form. The existence of analogous master
equations obtained by other approaches, but with different
expressions for $D_{pp}$ and $D_{xx}$ \cite{Mensky
2003,Petruccione-Breuer libro 2002}, has been described in
\cite{Barnett 2005}. The Caldeira-Leggett master equation
\cite{Caldeira-Leggett 1983}, is Markovian but is not of Lindblad
form and this has been shown to lead to serious difficulties
including, in particular, negative probabilities \cite{Ambegaokar
1991, Stenholm 1994, Barnett 2006}. These problems do not arise
for Eq. (\ref{starting master equation}) because of the presence
in $D_{pp}$ of two new terms other than the temperature dependent
one, $\gamma Mk_BT$. These depend on the variances $\Delta_\sigma
p$ and $\Delta_\sigma x$, and in particular the presence of a new
double commutator term representing a position diffusion regulated
by $D_{xx}$. The origin of these terms is the inherent spreading
in position(momentum) which occurs when a measurement of
momentum(position) is made. This point has been discussed further
in \cite{Barnett 2006}.

The main aim of this paper is an analysis of the decoherence
dynamics associated with Eq.\,(\ref{starting master equation}). In
particular we find the exact solution of the master equation, and
then use this to illuminate the role of both the $D_{pp}$ and the
hitherto largely unconsidered $D_{xx}$ terms in dissipation and
decoherence. To this end we will consider two initial states for
the Brownian particle: a single Gaussian wave packet and
Schr\"{o}dinger cat state. By way of comparison we also consider
cases where $D_{pp}$ and $D_{xx}$, can be varied independently,
which can, for example, furnish the Caldeira-Leggett dynamics when
$D_{pp}=\gamma Mk_BT$ and $D_{xx}=0$.

The paper is organized as follows. In Sec. \ref{par:Solving the
Master equation} we solve the master equation, providing a general
scheme for obtaining time evolution from general initial
conditions. In Sec. \ref{par:Applications} we apply this scheme to
two initial configurations, and discuss their dynamical evolution.
In Sec. \ref{par:Intermediate region} we consider briefly what
happens when the Lindblad condition given by Eq.\,(\ref{Lindblad
condition}) is not satisfied. In Sec. \ref{par:Conclusion} we
summarize and discuss our results. In \ref{par:solving the master
equation app} and \ref{Par:solving schrodinger cat dynamics} we
collect some of the lengthier calculations.

\section{Solution of the master equation \label{par:Solving the Master
equation}}

In this section we solve the master equation by introducing a
characteristic function. This procedure has been previously used
to obtain a formal solution of the Caldeira-Leggett master
equation, and is described, for example, in \cite{Libro decoerenza
2002}. The main difference here is that there is an additional
term depending on the position diffusion $D_{xx}$.

In the position representation Eq.\,(\ref{starting master
equation}) takes the form:
\begin{eqnarray}\label{}
 \!\!\!\!\!\!\!\!\!\!\!\!\!\!\!\!\! \frac{\partial \rho (x,x',t)}{\partial
t}  =&& \left[\frac{i \hbar}{2 M}\left(\frac{\partial^2}{\partial
x^2}-\frac{\partial^2}{\partial x'^2} \right)-
\frac{i\gamma}{2}(x-x')\left(\frac{\partial}{\partial
x}-\frac{\partial}{\partial x'} \right)-\frac{D_{pp}}{
\hbar^{2}}(x-x')^2 \right. \nonumber \\  && - \left.
D_{xx}\left(\frac{\partial^2}{\partial
x^2}+\frac{\partial^2}{\partial x'^2}-2\frac{\partial}{\partial
x}\frac{\partial}{\partial x'} \right)  \right]\rho(x,x',t)
 \, ,
\end{eqnarray}
where $\rho(x,x',t)= \langle x|\rho|x^\prime \rangle$. This second
order linear partial differential equation can be greatly
simplified by moving to a ($k, \Delta_t $) representation
\cite{Unruh 1989} based on introducing the characteristic function
$\rho(k,\Delta_t,t)$:
\begin{eqnarray}\label{moving to delta k}
\fl
\rho(k,\Delta_t,t)=\mathrm{tr}(\hat{D}\hat{\rho})=\mathrm{tr}(\exp
[i(k \hat{x}+ \Delta_t \hat{p})]\hat{\rho})
=\int_{-\infty}^{+\infty} \!\!\mathrm{d} x \, \mathrm{e}^{i k x}
\rho \left(x+\frac{\hbar \Delta_t}{2},x-\frac{\hbar \Delta_t}{2},t
\right).
\end{eqnarray}
In this new representation we obtain a first order partial
differential equation in the form
\begin{equation}\label{delta k differential equation}
\!\!\!\!\!\!\!\!\frac{\partial \rho(k,\Delta_t,t)}{\partial
t}=\left[\frac{k}{ M}\frac{\partial}{\partial \Delta_t}-\gamma
\Delta_t \frac{\partial}{\partial \Delta_t}-D_{pp}\Delta^2_t-
D_{xx}k^2 \right]\rho(k,\Delta_t,t)\,.
\end{equation}
In \ref{par:solving the master equation app} the method of
characteristics \cite{Barnett 1997} is used to solve this partial
differential equation exactly. The solution is
\begin{eqnarray}\label{evoluzione matrice densità}
\fl \rho(k,\Delta_t,t)=\rho\left(k,\Delta_t
(1-\Gamma)+\frac{k}{\gamma M}\Gamma,0\right) \times \nonumber \\
\fl \exp \left\{ \left[  - \left(D_{xx}+\frac{D_{pp}}{M^2
\gamma^2} \right)t +\frac{D_{pp}}{M^2 \gamma^3}
\left(\frac{\Gamma^2}{2}+\Gamma \right) \right]k^2 -\frac{D_{pp}
\Gamma^2}{M \gamma^2}k \Delta_t -\frac{D_{pp}  \Gamma
(2-\Gamma)}{2 \gamma}\Delta_t^2\,\right\}
   ,
\end{eqnarray}
where $\Delta_0=\Delta_t (1-\Gamma)+\frac{k}{\gamma M}\Gamma$ and
$\Gamma = 1-\exp (- \gamma t)$.

We can use Eq.\,(\ref{moving to delta k}) and its inverse to move
to and from the coordinate and $(k, \Delta_t)$ representations at
will, and Eq.\,(\ref{evoluzione matrice densità}) to obtain the
time evolution in the $(k, \Delta_t)$ representation. This allows
us to compute the time evolution of a general initial state in the
position representation by following the procedure:
\begin{equation}\label{protocol}
   \rho(x,x',0) \rightarrow \rho(k,\Delta_0,0) \rightarrow
\rho(k,\Delta_t,t) \rightarrow
   \rho(x,x',t)\,.
\end{equation}

In the next section we will use this procedure to solve the master
equation for two initial states which can be written in the
coordinate representation as a sum of exponential terms of the
form
\begin{eqnarray}\label{initial spatial density matrix}
\rho(x,x',0)=&& \exp \left[ -A_0(x-x')^2 - i B_0
(x-x')(x+x')\nonumber
\right. \\
&&\left.  - C_0 (x+x')^2 -i D_0(x-x') -E_0(x+x') -F_0 \right]\,.
\end{eqnarray}
We here apply the scheme outlined by Eq.\,(\ref{protocol}) to
obtain the time evolution for states of this kind. Using
Eqs.\,(\ref{moving to delta k}) and (\ref{evoluzione matrice
densità}) we find the characteristic function at time $t$ in the
$(k, \Delta_t)$ representation to be
\begin{equation}\label{delta k density matrix}
\rho(k,\Delta_t,t)= \exp \left[ -a_tk^2 - i b_t k \Delta_t
-c_t\Delta^2_t    - i d_tk   -i e_t\Delta_t
 -f_t\right]\,,
\end{equation}
where the various coefficients follow the time evolution given by
\begin{eqnarray}\label{c coefficients evolution}
a_t&=& a_0+b_0\frac{\Gamma}{M \gamma}+c_0\left(\frac{\Gamma}{M
\gamma}\right)^2+\left(D_{xx}+\frac{D_{pp}}{M^2 \gamma^2} \right)t
-\frac{D_{pp}}{M^2 \gamma^3} \left(\frac{\Gamma^2}{2}+\Gamma
\right) ,\nonumber \\
b_t&=&b_0 (1-\Gamma)+  c_0\frac{ 2\Gamma (1-\Gamma)}{M\gamma}
+\frac{D_{pp} \Gamma^2}{M \gamma^2} , \quad c_t= c_0 (1-\Gamma)^2+
\frac{D_{pp} \Gamma (2-\Gamma)}{2
\gamma}  , \nonumber  \\
d_t&=& d_0 + e_0  \frac{\Gamma }{M \gamma}, \quad e_t=e_0
(1-\Gamma) , \quad  f_t=f_0\,,
\end{eqnarray}
and where the relation between small and capital coefficients is
given by
\begin{eqnarray}\label{to delta k representation at t=0}
&&a_0= \frac{1}{16 C_0} , \,\,\, b_0=-\frac{B_0 }{4 C_0} \hbar
 , \,\,\, c_0=\frac{4 A_0 C_0+B_0^2}{4 C_0}\hbar^2  , \quad d_0=\frac{E_0
}{4 C_0} \nonumber \\
&&  e_0=\frac{2 C_0 D_0 -B_0 E_0 }{2 C_0} \hbar , \quad\exp
(-f_0)= \exp (-F_0)\exp \left(-\frac{E_0^2}{4 C_0}\right)
\frac{1}{2} \sqrt{\frac{\pi}{C_0}} \,.
\end{eqnarray}

Transformation back to the coordinate representation
Eq.\,(\ref{delta k density matrix}) provides the density matrix
\begin{eqnarray}\label{spatial density matrix}
\rho(x,x',t)= &&\exp \left[-  A_t(x-x')^2-  i B_t(x-x')(x+x')
\nonumber\right. \\
&&\left. -  C_t(x+x')^2- i D_t(x-x')- E_t(x+x')- F_t \right]\,,
\end{eqnarray}
with the following relationships
\begin{eqnarray}\label{to coordinate representation}
&&A_t= \frac{4a_t c_t- b_t^2}{4 \hbar^2 a_t} ,\quad B_t=-\frac{
b_t}{4 \hbar a_t} , \quad C_t=\frac{1}{16 a_t} , \quad D_t=
\frac{2a_te_t-b_td_t}{2 \hbar a_t}  , \nonumber \\
&& E_t= \frac{ d_t}{4 a_t}, \quad\exp (-F_t)= \frac{\exp
(-f_t)}{2\sqrt{\pi a_t}}\exp \left(-\frac{d_t^2}{4 a_t}\right)\,.
\end{eqnarray}
The inverse relations can also be found:
\begin{eqnarray}\label{to delta k representation}
&&a_t= \frac{1}{16 C_t} , \,\,\, b_t=-\frac{B_t }{4 C_t} \hbar
 , \,\,\, c_t=\frac{4 A_t C_t+B_t^2}{4 C_t}\hbar^2  , \quad d_t=\frac{E_t
}{4 C_t} \nonumber \\
&&  e_t=\frac{2 C_t D_t -B_t E_t }{2 C_t} \hbar , \quad\exp
(-f_t)= \exp (-F_t)\exp \left(-\frac{E_t^2}{4 C_t}\right)
\frac{1}{2} \sqrt{\frac{\pi}{C_t}} \,.
\end{eqnarray}

For density matrices which correspond to superpositions of states
centred on different positions, each of which is individually of
the form given by Eq.\,(\ref{initial spatial density matrix}), the
linearity of Eq.\,(\ref{delta k differential equation}) means that
it is still possible to follow the procedure described by
Eq.\,(\ref{protocol}) and so to compute the time evolution.

\section{Applications \label{par:Applications}}

In this section we consider two physically interesting initial
states, the simple Gaussian wavepacket and a superposition of two
such states, which forms a Schr\"odinger cat state. Using the
procedure described in the preceding section to compute their time
evolution, we focus on the role of the momentum and position
diffusion terms, proportional to $D_{pp} $ and $D_{xx}$
respectively. In particular many of our results will be expressed
in terms of these coefficients and will not depend on their
explicit form [Eqs.\,(\ref{dipp}) and (\ref{dixx})] in terms of
the physical parameters of the system. These results have a
general validity, therefore for any master equation with the same
form as Eq.\,(\ref{starting master equation}) irrespective of the
sizes of the diffusion terms. Thus our method does not just
provide the solution to Eq.\,(\ref{starting master equation}), but
also an infinity of other master equations whose diffusion terms
do not necessarily satisfy the Lindblad condition
(Eq.\,(\ref{Lindblad condition})). One particular example is the
Caldeira-Leggett master equation obtained by putting
$D_{pp}=\gamma M k_B T$ and $D_{xx}=0$.

\subsection{Single Gaussian wave packet}

Consider a minimum uncertainty Gaussian wave packet centered at
position $x_0=0$ and momentum $p_0$, with initial spreads in
position and momentum ${\Delta x}_0$ and ${\Delta p}_0$ which
satisfy the uncertainty principle ${\Delta x}_0 {\Delta
p}_0=\frac{\hbar}{2}$,
\begin{equation}\label{Gaussian wave packet}
\rho(x,x',0)= \frac{1}{\sqrt{2 \pi {\Delta x}_0^2}}\exp
\left[-\frac{(x-x')^2}{8{\Delta x}_0^2}- \frac{(x+x')^2}{8{\Delta
x}_0^2}+i \frac{p_0(x-x')}{\hbar} \right]\,.
\end{equation}
By comparing this initial reduced density matrix with
Eq.\,(\ref{initial spatial density matrix}) at $t= 0$ we can
identify the required coefficients as
\begin{equation}
\fl A_0=C_0=\frac{1}{8{\Delta x}_0^2}, \quad B_0=E_0=0, \quad
D_0=- \frac{p_0}{\hbar} , \quad \exp (-F_0)= \frac{1}{\sqrt{2 \pi
{\Delta x}_0^2}} \,.
\end{equation}
By using Eq.\,(\ref{moving to delta k}) we can obtain the
corresponding initial condition in the ($k, \Delta_t $)
representation, which has the exponential form of Eq.\,(\ref{delta
k density matrix}) for $t=0$ with
\begin{eqnarray}
a_0=\frac{{{\Delta x}_0}^2}{2},\quad b_0=0, \quad
c_0=\frac{{{\Delta p}_0}^2}{2},  \quad d_0=0,\quad e_0=- p_0,
\quad f_0=0\,.
\end{eqnarray}
>From Eq.\,(\ref{c coefficients evolution}) we next compute the
time evolution of the various coefficients, and then return to the
coordinate representation using Eqs.\,(\ref{spatial density
matrix}) and (\ref{to coordinate representation}). Thus we obtain
the spatial reduced density matrix at time $t$, from which it is
also possible to obtain the corresponding density matrix in the
momentum representation by double Fourier transformation. The
solution provides a simple means of obtaining the time evolution
of the average of $\hat{x}$ and $\hat{p}$, and of their variances
shown below and plotted in Fig.\,1 for physically reasonable
parameter values,
\begin{eqnarray}\label{medie}
 \langle \hat{x}\rangle_t &=&-d_t, \,\,\, \langle
\hat{p}\rangle_t=-e_t, \,\,\, {\Delta x}^2_t=2a_t, \,\,\,  {\Delta
p}^2_t = 2c_t \, .
\end{eqnarray}
These equations could equally well be rewritten in terms of the
capitalized coefficients using Eq.\,(\ref{to delta k
representation}), but the evolution is given most simply in terms
of the initial conditions using Eq.\,(\ref{c coefficients
evolution}). The expectation values become
\begin{eqnarray}\label{media di x e p}
&&  \fl \langle \hat{x}\rangle_t
=\frac{p_0}{M}\frac{1-\exp(-\gamma t)}{\gamma} , \quad \langle
\hat{p}\rangle_t= p_0 \exp(-\gamma t), \quad {\Delta x}^2_t
={\Delta x}^2_0+\frac{{\Delta
p}^2_0}{M^2}\left[\frac{1-\exp(-\gamma t)}{ \gamma}\right]^2
 +  \nonumber
\\ &&  \fl  2\left(D_{xx}  +\frac{D_{pp}}{M^2 \gamma^2} \right)t
-\frac{D_{pp}}{M^2 \gamma^3} \left[\frac{(1-\exp(-\gamma
t))^2}{2}+1-\exp(-\gamma t) \right]  \nonumber
\\
&& \fl{\Delta p}^2_t = {\Delta p}^2_0\exp(-2\gamma
t)+D_{pp}\frac{[1-\exp(-2\gamma t)]}{\gamma}\, .
\end{eqnarray}
As would be expected, the diffusions $D_{pp}$ and $D_{xx}$ do not
affect the mean values of the position and momentum. Of particular
interest, however, is the evolution of the variances, whose
dependence on $D_{pp}$
 and $D_{xx}$, can be simply found from Eq.\,(\ref{media di x e p}) for
both short times $(t\ll \gamma^{-1})$
\begin{eqnarray}\label{variances1}
{\Delta x}^2_t \approx {\Delta x}^2_0+ 2 D_{xx} t , \qquad {\Delta
p}^2_t&\approx &{\Delta p}_0^2 +2 \left( D_{pp} - \gamma {\Delta
p}^2_0  \right)t\, ,
\end{eqnarray}
and long times $(t\gg \gamma^{-1})$
\begin{eqnarray}\label{variances2}
{\Delta x}^2_t \approx  2\left( D_{xx} + \frac{D_{pp}}{M^2
\gamma^2} \right) t , \qquad {\Delta p}^2_t &\approx &
\frac{D_{pp}}{\gamma}\,.
\end{eqnarray}
Eqs.\,(\ref{variances1}) and (\ref{variances2}) reveal the
critical role of $D_{pp}$ and $D_{xx}$. In particular, for small
times ${\Delta x}^2_t$ depends only on $D_{xx}$, which shows that
the $D_{xx}$ term has an important role in this time region.
\begin{figure}[h!]\label{fig media di x e p}
\begin{center}
\includegraphics[width=3.8 cm, height=3 cm]{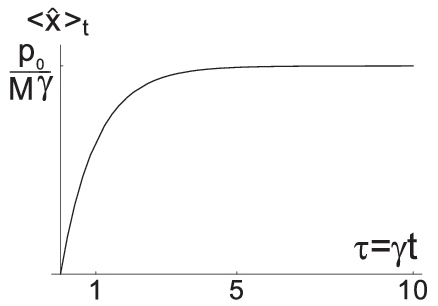}
\includegraphics[width=3.8 cm, height=3 cm]{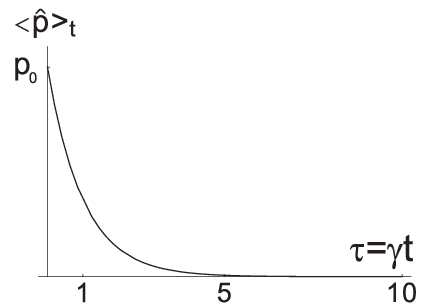}
\includegraphics[width=3.6 cm, height=3 cm]{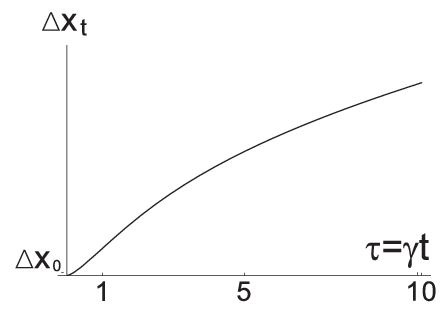}
\includegraphics[width=4 cm, height=3 cm]{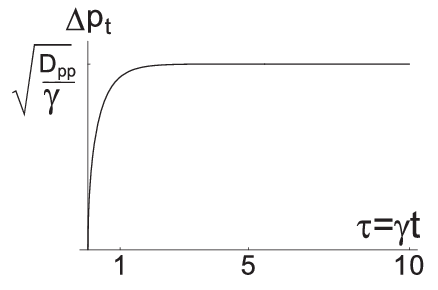}
\caption{\footnotesize{Figures on the left: evolution of the
average of $\hat{x}$ and $\hat{p}$. Figures on the right:
evolution of the variances ${\Delta x}_t$} and ${\Delta p}_t$.
Parameter values (in SI units):
 $M=5.01\times 10^{-22}, \quad m=5.01\times 10^{-26}, \quad
k_B=1.38\times 10^{-23}, \quad \hbar= 1.06\times 10^{-34},
 \quad \gamma=1000, \quad R = \gamma M/2 m= 5\times 10^{6},  \quad  T=300
K, \quad p_0= 5.01\times 10^{-26},
 \quad {\Delta x}_0=0.73 \times 10^{-7}, \quad {\Delta p}_0=7.26 \times
10^{-28}  \quad
 (\Delta_\sigma p)^2=(\Delta_\sigma x)^2=\frac{\hbar}{2}\times
n,\quad n=10^{4}$. These parameters remain the same in all the
following figures.}
\end{center}
\end{figure}

It is possible to calculate the evolution of the purity,
$\mathrm{tr}(\hat{\rho}^2)_t$, of the initial state of
Eq.\,(\ref{Gaussian wave packet}). This quantity, bounded by 0 and
1, is related to the linear entropy and is equal to 1 for pure
states. Any difference from 1 means a loss of purity of the state.
The purity of the Gaussian state is plotted in Fig.\,2 as a
function of time.
\begin{figure}[h!]\begin{center}
\label{traccia di ro}
\includegraphics[width=6.4 cm, height=4.4 cm]{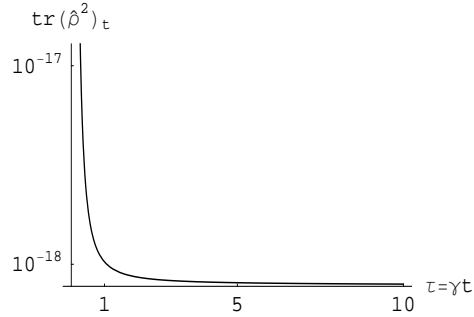}
\caption{\footnotesize{Behavior in time of
$\mathrm{tr}(\hat{\rho}^2)_t$.}} \end{center}
\end{figure}
The figure shows that the decoherence process occurs on a time
scale much smaller than the relaxation time of the particle, which
is of the order of $\gamma^{-1}$. The dynamics of our system are
then described by the particle density matrix time evolution as a
rapid transformation from the pure initial state (\ref{Gaussian
wave packet}) into a statistical mixture, as is shown in Fig.\,3
in the coordinate representation.
\begin{figure}[h!]\begin{center}
\label{plot of particle density matrix}
\includegraphics[width=7.2 cm, height=4.65 cm]{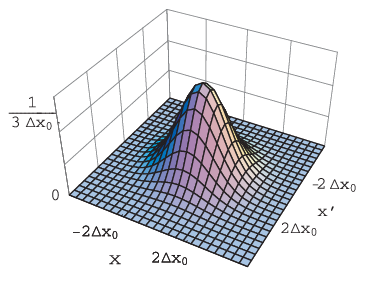}
\includegraphics[width=8 cm, height=4.95 cm]{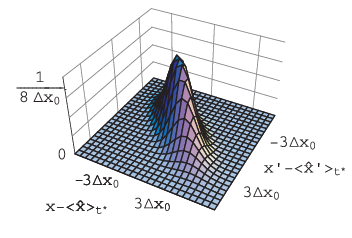}
\caption{\footnotesize{Left: Absolute value of the initial density
matrix (\ref{Gaussian wave packet}). Right: $|\rho(x,x',t^*)|$,
where $t^*\ll \gamma^{-1}$. The suppression of the off-diagonal
terms reflects the onset of decoherence before significant
spreading has occurred.}} \end{center}
\end{figure}

To investigate further the dynamics of this loss of coherence both
in $x$ and $p$ space we can compute the spreads
$\mathcal{M}_x^{2}(t)$ and $\mathcal{M}_p^{2}(t)$, and coherence
lengths, $\mathcal{L}_x(t)$ and $\mathcal{L}_p(t)$, using their
general definition in \cite{Barnett 2001} in terms of traces over
the density matrix of the particle
\begin{equation}\label{amplitude definition}
    \mathcal{M}_x^{2}(t)=\frac{\mathrm{Tr}(\hat{\rho}^{2}\hat{x}^{2})
    +\mathrm{Tr}(\hat{\rho}\hat{x}\hat{\rho}\hat{x})}{\mathrm{Tr}(\hat{\rho}^{2})}
    -
2\left(\frac{\mathrm{Tr}(\hat{\rho}^{2}\hat{x})}{\mathrm{Tr}(\hat{\rho}^{2})}\right)^2,
\end{equation}
\begin{equation}\label{coherence length definition}
    \mathcal{L}_x^{2}(t)=\frac{\mathrm{Tr}(\hat{\rho}^{2}\hat{x}^{2})
    -\mathrm{Tr}(\hat{\rho}\hat{x}\hat{\rho}\hat{x})}{\mathrm{Tr}(\hat{\rho}^{2})}\,,
\end{equation}
with similar expressions for the momentum spread and coherence
length. For pure states $\hat{\rho}^{2}=\hat{\rho}$ and therefore
both spread and coherence length become equal to the respective
width of the state, e.g.  $\mathcal{M}_x(t)=
\mathcal{L}_x(t)=\Delta x_t$. However, in presence of the
interaction the state of the particle loses its initial purity and
the two quantities differ. While $\mathcal{M}_x(t)$ gives the
extension of the state, $\mathcal{L}_x(t)$ gives the zone inside
the state extension, where coherence has not yet been lost at time
$t$ \cite{Libro decoerenza 2002}.

For an initial Gaussian wave packet the spread $\mathcal{M}_x (t)$
corresponds to the width of the reduced density matrix along the
main diagonal, $\mathcal{M}_x (t)=\Delta x_t =\int \rho(x,x,t)
x^2-(\int \rho(x,x,t) x)^2$, while the coherence length
$\mathcal{L}_x(t)$ gives analogously the width of the reduced
density matrix along the main skew diagonal $\mathcal{L}_x(t)=\int
\rho(x,-x,t) x^2-(\int \rho(x,-x,t) x)^2 $.

The ratio $\mathcal{L}_x(t)/\mathcal{M}_x(t)$ gives a
dimensionless measurement of the loss of coherence. It is
interesting to note that for an initial Gaussian wave packet
\cite{Morikawa 1990}, this ratio is directly connected to
$\mathrm{tr}(\hat{\rho}^2)$ and is equal in both the position and
momentum representations. This property is found also in our
system where the following relations are obtained
\begin{equation}\label{adimensional ratio}
\mathrm{tr}(\hat{\rho}^2)_t=1-S_{lin}=
\frac{\mathcal{L}_x(t)}{\Delta x_t}=\frac{\mathcal{L}_p(t)}{\Delta
p_t}=\frac{\hbar /2}{\sqrt{4a_tc_t- b_t^2}} \,  .
\end{equation}
>From Eqs.\,(\ref{adimensional ratio}) and (\ref{medie}) we obtain
\begin{equation}\label{diseguaglianza generalizzata}
   \mathcal{L}_x(t) \Delta p_t =  \mathcal{L}_p(t) \Delta x_t
   =\frac{\hbar}{2}\sqrt{\frac{1}{1-b^2_t/4a_tc_t}}\geq\frac{\hbar}{2},
\end{equation}
which can be seen as a particular case of the generalized
uncertainty relation derived in \cite{Barnett 2001}.

The squares of the coherence lengths in the two representations
are given by
\begin{equation}\label{lunghezza di coerenza spaziale e nei momenti}
\mathcal{L}_x^2(t)= \frac{\hbar^2 a_t/2} {4a_tc_t-b_t^2} \quad
\mathrm{and}
 \quad \mathcal{L}_p^2(t)= \frac{\hbar^2 c_t/2} {4a_tc_t-b_t^2} \,.
\end{equation}
It is instructive to consider the evolution of the coherence
lengths at short times $(t\ll \gamma^{-1})$
\begin{eqnarray}\label{lunghezza di coerenza spaziale e nei momenti small
times} \mathcal{L}_x^2(t)\approx \Delta x_0^2\left[1- 2
\left(\frac{D_{pp}}{\Delta p_0^2} - \gamma
 \right) t \right], \qquad
\mathcal{L}_p^2(t)\approx  \Delta p_0^2\left(1 - 2
\frac{D_{xx}}{\Delta x_0^2}t \right) \,,
\end{eqnarray}
and for large times $(t\gg \gamma^{-1})$
\begin{eqnarray}\label{lunghezza di coerenza spaziale e nei momenti large
times} \mathcal{L}_x^2(t)\approx  \frac{\hbar^2}{4 \Delta
p_t^2}\approx \frac{\hbar^2 \gamma }{4 D_{pp}}, \qquad
\mathcal{L}_p^2(t)\approx \frac{\hbar^2}{4 \Delta x_t^2} \,.
\end{eqnarray}
Eqs.\,(\ref{lunghezza di coerenza spaziale e nei momenti small
times}) and (\ref{lunghezza di coerenza spaziale e nei momenti
large times}) show the role of $D_{pp}$ and $D_{xx}$ in these
evolutions, i.e. in the decoherence process. In particular, for
small times the momentum coherence length $\mathcal{L}_p^2(t)$
depends only on the position diffusion $D_{xx}$, showing again the
relevance of this term in this time region. This behaviour of
$\mathcal{L}_p^2(t)$ can be understood by comparing it with that
of $\mathcal{L}_x^2(t)$ which, as it is independent of $D_{xx}$,
holds also in the Caldeira-Leggett model. Indeed, from
Eqs.\,(\ref{lunghezza di coerenza spaziale e nei momenti small
times}) and (\ref{variances1}) we see that, as
$\mathcal{L}_x^2(t)$ and $\Delta p^2_t$ both depend on the factor
$\left(D_{pp}/\Delta p_0^2 - \gamma \right)$, and the same occurs
for $\mathcal{L}_p^2(t)$ and $\Delta x^2_t$, which both depend on
$D_{xx}$.

\subsection{Schr\"{o}dinger cat state}

The second configuration that we investigate is an initial
Schr\"{o}dinger cat state with a model wavefunction of the form
\begin{eqnarray}\label{Schrodinger cat}
&&\psi (x,0) = \frac{1}{\sqrt{2\left[1+\exp
\left(-\frac{l^2}{8\sigma^2}\right)\right]} (2\pi
\sigma^2)^{-\frac{1}{4}} } \times \nonumber
\\
&& \left\{ \! \exp \! \left[\!
-\frac{\left(x-\frac{l}{2}\right)^2}{4\sigma^2}-i \frac{M
v}{\hbar}x\right]\! + \!\exp \! \left[ \!
-\frac{\left(x+\frac{l}{2}\right)^2}{4\sigma^2}+i \frac{M
v}{\hbar}x \! \right] \!\right\}\ ,
\end{eqnarray}
where $\sigma$ is the width of two wave packets initially placed
at a distance $l$ with one moving towards the other with an
initial velocity $v$. Such states are interesting because of their
potentially long-range coherence properties and the extreme
sensitivity of this to environmental decoherence \cite{Brune
1997}.

In the absence of any interactions the behaviour of the diagonal
reduced density matrix elements is given by
\begin{eqnarray}\label{schrodinger cat final density matrix free case}
&& \fl \rho(x,x,t)=\frac{1}{2\left(1+\exp
[-\frac{l^2}{8\sigma^2}]\right) \sqrt{2 \pi
\left(\sigma^2+\frac{\hbar^2 t^2}{4 M^2 \sigma^2}\right)}}
  \left\{ \exp \left[ -
\frac{(x-\frac{l}{2}+vt)^2}{2\left(\sigma^2+\frac{\hbar^2 t^2}{4
M^2 \sigma^2}\right)} \right] +   \nonumber \right. \\ &&
 \left.  \fl\exp \left[  -
\frac{(x+\frac{l}{2}-vt)^2}{2\left(\sigma^2+\frac{\hbar^2 t^2}{4
M^2 \sigma^2}\right)}  \right] + \exp \left[ - \frac{x^2
+\left(\frac{l}{2}-vt\right)^2}{2\left(\sigma^2+\frac{\hbar^2
t^2}{4 M^2 \sigma^2}\right)} \right] \cos \left[\frac{\frac{4 M
v\sigma^2}{\hbar}+\frac{\hbar l t}{2 \sigma^2
M}}{2\left(\sigma^2+\frac{\hbar^2 t^2}{4 M^2 \sigma^2}\right)}
x\right] \right\}  \,.
\end{eqnarray}
This probability distribution is the sum of three contributions.
The first two clearly correspond to a pair of separately expanding
(freely spreading) wave packets, while the third term represents
an interference term, responsible for the central peak present in
Fig.\,4, which is a plot of Eq.\,(\ref{schrodinger cat final
density matrix free case}). This is exactly as would be expected
for such undamped evolution.
\begin{figure}[h]\label{Schrodingercatfreecase}
\begin{center}
\includegraphics[width=6.8 cm, height=4.8 cm]{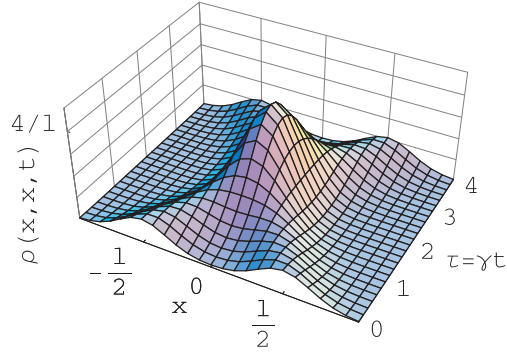}
\caption{\footnotesize{Free evolution of the diagonal reduced
density matrix elements of Eq.\,(\ref{schrodinger cat final
density matrix free case})}. Used values (SI): $l=4\times 10^{-7},
\sigma =0.73 \times 10^{-7}, v=10^{-4}$.}
\end{center}\end{figure}

In \ref{Par:solving schrodinger cat dynamics} the time evolution
is computed from the initial state of Eq.\,(\ref{Schrodinger cat})
using both the procedure described in Eq.\,(\ref{protocol}) and
the linearity of Eq.\,(\ref{delta k differential equation}). Along
the $x=x^\prime$ diagonal we find
\begin{eqnarray}\label{schrodinger cat final density matrix^2}
& & \fl \rho (x,x,t)=\frac{1}{2\left[1+\exp
(-\frac{l^2}{8\sigma^2})\right] 2 \sqrt{\pi \bar{a}_t}}   \left\{
\exp \left[  - \frac{(x+\bar{d}_t)^2}{4 \bar{a}_t}\right]+\exp
\left[ - \frac{(x-\bar{d}_t)^2}{4 \bar{a}_t} \right] \nonumber
\right.
\\ && \left.
 \!\!\!\! + \exp  \! \left(- \frac{{x^2
-|\bar{d}_t|}^2}{4 \bar{a}_t} -\frac{2 M^2 v^2
\sigma^2}{\hbar^2}-\frac{l^2}{8 \sigma^2} \right) \cos
\frac{|\bar{d}_t| x}{2 \bar{a}_t} \right\},
\end{eqnarray}
with the evolution of the various coefficients given in
Eq.\,(\ref{evolution of coefficients}). Eq.\,(\ref{schrodinger cat
final density matrix^2}) reduces to Eq.\,(\ref{schrodinger cat
final density matrix free case}) in the absence of any
interaction, as may readily be verified by substituting
$\lim_{\gamma\rightarrow 0}\Gamma/\gamma = t$ and
$D_{pp}=D_{xx}=0$. This probability distribution is again the sum
of three contributions. In order to investigate the behavior of
the interference term, we compute the attenuation coefficient
$W_t$ \cite{Ford 2001}, defined as the ratio of the factor
multiplying the cosine interference term to twice the geometric
mean of the first two terms:
\begin{equation}\label{attenuation coefficient}
W_t=\exp \left[ \frac{{\bar{d}_t}^2+|\tilde{d}_t|^2}{4
a_t}-\frac{l^2}{8\sigma^2} - \frac{2 M^2 v^2 \sigma^2}{\hbar^2}
\right] \,.
\end{equation}
In the free case $W_t=1$ for all times, corresponding to a full
coherence. In the presence of interaction $W_t$ decays quickly, as
pictured in Fig.\,5, which shows the rapid destruction of the
interference with time.
\begin{figure}[h!]
\begin{center}\label{plot of attenuation coefficient}
\includegraphics[width=6.3 cm, height=4.8 cm]{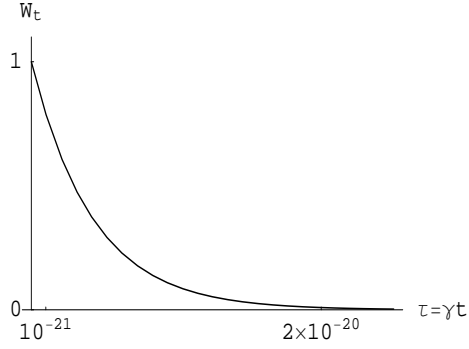}
\caption{\footnotesize{Evolution of the attenuation coefficient
$W_t$.}}
\end{center}\end{figure}

By expanding $W_t$, for small times, in a Taylor series we find:
\begin{equation}\label{attenuation coefficient for small times}
W_t \approx 1-D_{xx}\left(\frac{l^2}{4\sigma^2} + \frac{4 M^2 v^2
\sigma^2}{\hbar^2}\right) t   \,.
\end{equation}
This equation shows that for very short times the attenuation
factor decays with a characteristic time $\tau_D$ given by
\begin{equation}\label{decoherence time}
\tau_D= \left[D_{xx}\left(\frac{l^2}{4\sigma^2} + \frac{4 M^2 v^2
\sigma^2}{\hbar^2}\right)\right]^{-1}   \,.
\end{equation}
The interaction with the environment leads to the destruction of
the interference term. The initial decay of $W_t$, which
characterizes this, is due solely to the presence of the $D_{xx}$
term in Eq.\,(\ref{starting master equation}). All of the
decoherence occurs on a very short timescale and is due entirely
to this diffusion.

In order to investigate further the dynamics of this loss of
coherence in $x$ and $p$ space we compute the spreads,
$\mathcal{M}_x^{2}(t)$ and $\mathcal{M}_p^{2}(t)$, and coherence
lengths, $\mathcal{L}_x(t)$ and $\mathcal{L}_p(t)$ defined in
Eqs.\,(\ref{amplitude definition}) and (\ref{coherence length
definition}). By using Eq.\,(\ref{schrodinger cat final total
density matrix^2}) we obtain the spatial spread and the coherence
length at time $t$
\begin{eqnarray}\label{width for cat}
   &&  \fl \mathcal{M}_x^{2}(t)=\frac{1}{ 8 \bar{C}_t}+\frac{1}{
8\bar{C}_t^2}\left\{\bar{E}_t^2
   \exp \left(\frac{\bar{E}_t^2}{2\bar{C}_t}-2\tilde{F}_t\right)
     - \tilde{E}_t^2 \exp
    \left(-\frac{\tilde{E}_t^2}{2\bar{C}_t}-2\bar{F}_t\right)+
     \right.  \\
    &&  \!\!\!\! \left.\left[\left(\bar{E}_t^2-\tilde{E}_t^2\right) \cos
\alpha - 2\bar{E}_t\tilde{E}_t \sin \alpha \right] \exp
    \left(-\beta\right) \rule{0cm}{0.5cm}\right\} \div   \nonumber \\
    &&\fl
    \left\{
    \exp \left(\frac{\bar{E}_t^2}{2\bar{A}_t} -2\bar{F}_t\right)
    +\exp \left(\frac{\tilde{D}_t^2}{2\bar{A}_t}-2\tilde{F}_t\right)+
    \exp \left(-\frac{\bar{D}_t^2}{2\bar{A}_t}-2\bar{F}_t\right)   +\exp
    \left(-\frac{\tilde{E}_t^2}{2\bar{A}_t}-2\tilde{F}_t\right)\right\}\,,
\nonumber
\end{eqnarray}
and
\begin{eqnarray}\label{coherence length for cat}
   &&\fl  \mathcal{L}_x^{2}(t)=\frac{1}{ 8 \bar{A}_t}+\frac{1}{
8\bar{A}_t^2}\left\{\tilde{D}_t^2
   \exp \left(\frac{\tilde{D}_t^2}{2\bar{A}_t}-2\tilde{F}_t\right)
     - {\bar{D}_t}^2 \exp
    \left(-\frac{{\bar{D}_t}^2}{2\bar{A}_t}-2\bar{F}_t\right)+
     \right.  \\
    && \left.\left[\left(\tilde{D}_t^2-\bar{D}_t^2\right) \cos \alpha
- 2\bar{D}_t\tilde{D}_t \sin \alpha \right] \exp
    \left(-\beta\right) \rule{0cm}{0.5cm}\right\} \div   \nonumber \\
    &&\fl
    \left\{
    \exp \left(\frac{\bar{E}_t^2}{2\bar{A}_t} -2\bar{F}_t\right)
    +\exp \left(\frac{\tilde{D}_t^2}{2\bar{A}_t}-2\tilde{F}_t\right)+
    \exp \left(-\frac{\bar{D}_t^2}{2\bar{A}_t}-2\bar{F}_t\right)   +\exp
    \left(-\frac{\tilde{E}_t^2}{2\bar{A}_t}-2\tilde{F}_t\right)\right\}\,,
\nonumber
\end{eqnarray}
where $\alpha
=\left(\bar{C}_t\bar{D}_t\tilde{D}_t+\bar{A}_t\bar{E}_t\tilde{E}_t\right)
/4\bar{A}_t\bar{C}_t+\bar{F}_t+\tilde{F}_t$ and
$\beta=\left[\bar{C}_t\bar{D}_t^2-\bar{C}_t\tilde{D}_t^2+A\tilde{E}_t^2-\bar{A}_t\bar{E}_t^2\right]$
$/8\bar{A}_t\bar{C}_t$. In order to obtain the same quantities in
momentum space we need the corresponding reduced density matrix.
If we perform a double Fourier transform we obtain $\rho(p,p',t)$
from $\rho(x,x',t)$. The result has the same form as
Eq.\,(\ref{schrodinger cat final total density matrix^2}) with the
following substitutions:
\begin{eqnarray}\label{passaggio da x a p}
   \fl  \bar{A}_t \!\! \rightarrow  \!\!
    \frac{\bar{A}_t}{Z_t},\quad
    \bar{B}_t  \rightarrow
    \frac{-\bar{B}_t}{\bar{Z}_t},\quad \bar{C}_t  \rightarrow
    \frac{\bar{C}_t}{Z_t},
    \quad
     \bar{D}_t \!\!  \rightarrow  \!\!  2 \hbar
    \frac{\bar{B}_t \bar{D}_t + 2\bar{A}_t \bar{E}_t }{Z_t},\quad
    \tilde{D}_t  \rightarrow  - 2 \hbar
    \frac{\bar{B}_t \tilde{D}_t + 2 \bar{A}_t \tilde{E}_t }{Z_t},
    \nonumber \\ \fl
     \bar{E}_t \!\!   \rightarrow   \!\!  2 \hbar
    \frac{\bar{B}_t \bar{E}_t -2 \bar{C}_t \bar{D}_t }{Z_t}, \quad
    \tilde{E}_t  \rightarrow   2 \hbar
    \frac{\bar{B}_t \tilde{E}_t-2 \bar{C}_t \tilde{D}_t}{Z_t},
    \quad
     \bar{F}_t \!\!   \rightarrow \!\! \bar{F}_t \!\! + \frac{\ln  \!
Z_t}{2} + \nonumber \\ \fl 4  \hbar^2
     \frac{\bar{C}_t \bar{D}_t^2 - \bar{B}_t \bar{D}_t\bar{E}_t - \bar{A}_t
     \bar{E}_t^2}{Z_t},\quad
     \tilde{F}_t \!\!   \rightarrow   \!\! \tilde{F}_t \!\! + \frac{\ln
\!  Z_t}{2}  +4 \hbar^2
     \frac{\bar{C}_t \tilde{D}_t^2 - \bar{B}_t \tilde{D}_t\tilde{E}_t -
\bar{A}_t
     \tilde{E}_t^2}{Z_t} ,
\end{eqnarray}
where $Z_t=4(\bar{B}_t^2+4\bar{A}_t\bar{C}_t)\hbar^2$. By using
these substitutions in Eqs.\,(\ref{width for cat}) and
(\ref{coherence length for
 cat}) one obtains the corresponding $\mathcal{M}_p^{2}(t)$ and
 $\mathcal{L}_p^{2}(t)$.

These quantities satisfy
 $\mathcal{M}_x(0)=\mathcal{L}_x(0)$ and
 $\mathcal{M}_p(0)=\mathcal{L}_p(0)$ at the initial time, while for large
times their increase is the same as the corresponding quantities
$\Delta x_t$, $l_x (t)$, $\Delta p_t$ and $l_p (t)$ in
 Eqs.\,( \ref{variances2}) and (\ref{lunghezza di coerenza spaziale e nei
momenti large
 times}) found for the single Gaussian wave packet. In particular it has
been shown that a generalized uncertainty relation holds
\cite{Barnett 2001}
\begin{equation}\label{uncertainty relationship}
     \mathcal{L}_x(t)\mathcal{M}_p(t) \geq
     \frac{\hbar}{2}\,.
\end{equation}
This product is plotted in Fig.\,6 for our cat state, which shows
how the uncertainty relation is satisfied at all times.
\begin{figure}[h!]
\begin{center}
\includegraphics[width=6.3 cm, height=4.8 cm]{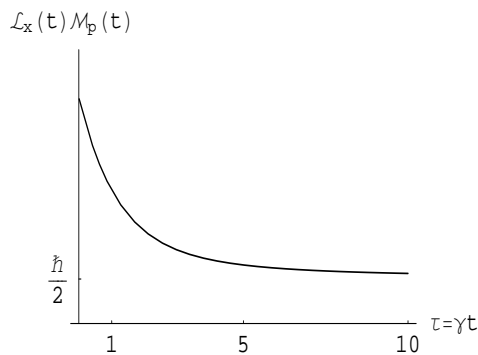}
\caption{\footnotesize{The evolution of the product
$\mathcal{L}_x(t)\mathcal{M}_{p}(t)$.}}
\end{center}\end{figure}

\section{Unphysical parameter region \label{par:Intermediate region}}

As was stated in the introduction, one of the key differences
between Eq.\,(\ref{starting master equation}) and the Caldeira
Leggett equation is the presence of the $D_{xx}$ term. The latter
master equation, of course, is not of the Lindblad type and
pathological behavior has been observed by several authors
\cite{Libro decoerenza 2002,Barnett 2006,Stenholm 1994} if
$\gamma$ is too large or the wave packet width too small (smaller
than the thermal de Broglie wavelength of the object). If we
consider the time derivative of the linear entropy for small
times, then we would expect a positive value for an initial pure
state, for which $S_{lin}=0$. In our model we find this time
derivative to be
\begin{eqnarray}\label{derivative of linear entropy}
\fl \frac{\mathrm{d}}{\mathrm{d}
t}S_{lin}|_{t=0}=-\frac{\mathrm{d}}{\mathrm{d} t}\mathrm{tr}
\hat{\rho}^2|_{t=0} =-2\langle \dot{\hat{\rho}} \rangle
=\frac{4}{\hbar^2}{{\Delta p}_0}^2 D_{xx}+
\frac{4}{\hbar^2}{{\Delta x}_0}^2 D_{pp} - \gamma \,,
\end{eqnarray}
where the brackets indicate the average over a general state of
the Brownian particle and we have used Eq.\,(\ref{starting master
equation}) to obtain the last equality. If we use ${\Delta
x}_0{\Delta p}_0=\frac{\hbar}{2}$, then the positivity of
Eq.\,(\ref{derivative of linear entropy}) is assured for all
possible states only if
\begin{equation}\label{Lindblad condition2}
D_{xx}D_{pp}\geq \left(\frac{\hbar \gamma}{4} \right)^2\, .
\end{equation}
This is the same condition as that found in \cite{Barnett 2006} by
requiring an initial reduction in the probability of remaining in
the initial pure state.

Even if in our model this condition is satisfied by our parameters
of Eqs.\,(\ref{dipp}) and (\ref{dixx}) it is interesting to work
near the region of its validity. In fact we can use our master
equation Eq.\,(\ref{starting master equation}), and look for
anomalous features when Eq.\,(\ref{Lindblad condition2}) is not
satisfied. For example, after substituting $D_{pp}= q
\frac{\left(\hbar \gamma / 4 \right)^2}{D_{xx}}$, we can vary $q$
around one. In Fig.\,7 Eq.\,(\ref{derivative of linear entropy})
is plotted as a function of $q$.
\begin{figure}[h]\label{Derivataentropialineare}
\begin{center}
\includegraphics[width=6.8 cm, height=4.8 cm]{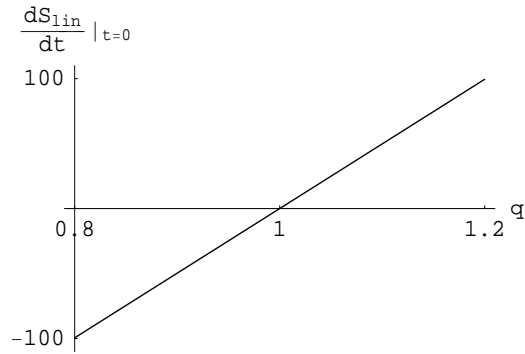}
\caption{\footnotesize{Behavior of the derivative of the linear
entropy at $t=0$ as a function of $q$.}}
\end{center}
\end{figure}

This figure clearly shows that the rate of change of linear
entropy is negative at small times if the Lindblad condition is
not satisfied.

\section{Conclusions and discussion \label{par:Conclusion}}

In this paper we have used the Lindblad master equation for
Brownian motion found in \cite{Barnett 2005, Barnett 2006} to
analyze wave packet dynamics. To this end we have provided a
simple and clear scheme by which we can obtain the exact time
evolution starting from general initial conditions.

We have applied this procedure to find the time evolution of two
initial states: the first in which the Brownian particle is
represented by a Gaussian wave packet and the second, in which it
is represented by a Schr\"{o}dinger cat state. In each case we
have provided expressions for the relevant quantities in both the
position and momentum representations. There are complementary
aspects of the two representations, such as generalized
uncertainty relations not found when one focuses only on the
spatial features.

We have analyzed further the dissipation and decoherence, in
particular focusing on the reduced density matrix evolution in the
coordinate representation. We have obtained the evolution of the
wavepacket widths and the coherence lengths for both initial
conditions and in both the position and momentum representations.
The Gaussian wave packet shows a very rapid decoherence on a
timescale much shorter than the relaxation time of the system. In
the momentum representation this decoherence depends only on the
position diffusion coefficient, $D_{xx}$, present in the master
equation. The $D_{xx}$ term is also the only term responsible for
the increase in the spatial width of the wave packet for small
times. The term containing this coefficient is absent in many
previous master equations which have been used to describe
friction and Brownian motion.

The evolution of a Schr\"odinger cat state without any frictional
effects shows coherent oscillations caused by interference between
the two components of the wave function. Our analysis shows that
such interference is damped on timescales again much shorter than
the relaxation time; decoherence is very rapid. We have
quantified the decoherence using an attenuation coefficient for
the oscillatory terms. This attenuation coefficient also depends
on $D_{xx}$. The results of the computation of generalized
variances and coherence lengths show that for large times these
quantities behave in a similar way to the single
Gaussian wave packet case, and that a generalized uncertainty relation
between variances and coherence lengths holds for all times.

Finally we have generalized the system to look at the case where
the product of the diffusion coefficients is not large enough to guarantee that
the master equation is of Lindblad form. Here we see a clear
signature of unphysical behaviour in the linear entropy, which is
associated with probabilities outside the physically-meaningful
range between 0 and 1.

The measurement-based quantum description of friction illustrated
in this paper provides a general framework for investigating the
role that the various terms in the master equation play in
decoherence. It is clear from our analysis that the two extra
diffusion terms not associated with the temperature of the system
are necessary to ensure complete positivity of the density
operator at all times. This is consistent with previous work
\cite{Barnett 2005,Barnett 2006}. The minimum sizes of these terms
are governed by an uncertainty relation, in line with their wholly
quantum origin. Diffusion in one observable is associated with
localization in the complementary one. These localizations occur
each time a measurement is made. The consequent diffusion, and in
particular that of position, which has no analogue in the
Caldeira-Leggett reservoir-based approach, must be taken account of
in any complete quantum description of friction. The cost of not
doing so is illustrated emphatically here; the resultant
incomplete equation cannot describe decoherence correctly, because
it is not valid on the short timescales during which decoherence
occurs.

The generality of the measurement-based approach reflects the
generality of the Kraus formalism of quantum measurements on which
it is based, which makes no reference to any particular
measurement device. Consequently the theory presented here is not
specific to any particular frictional or Brownian system. Such a
linkage could in principle be found for particular systems, and
would amount to an ab initio quantum theory of friction. No such
theory is known at this time.

\ack

We would like to thank Professor G. Compagno for useful
discussions.

\appendix

\section{}\label{par:solving the master equation app}
In this appendix we use the method of characteristics
\cite{Barnett 1997} to solve the master equation of
Eq.\,(\ref{delta k differential equation}). The first step is to
rewrite Eq.\,(\ref{delta k differential equation}) as
\begin{eqnarray}\label{delta k differential equation method of
characteristics form} \fl 1 \frac{\partial
\rho(k,\Delta_t,t)}{\partial t}+ \left(\gamma \Delta_t - \frac{k}{
M} \right) \frac{\partial \rho(k,\Delta_t,t) }{\partial \Delta_t}
= -\left(D_{pp}\Delta_t^2 + D_{xx}k^2 \right) \rho(k,\Delta_t,t)
\,.
\end{eqnarray}
The curves in the $t,\Delta_t$ plane, parameterized by $l$ and
defined by the relation
\begin{equation}\label{characteristics curves}
 \frac{\mathrm{d} t}{1}= \frac{\mathrm{d} \Delta_t}{\gamma \Delta_t -
\frac{k}{ M}}=\mathrm{d}
 l\, ,
\end{equation}
are called the characteristic curves of the partial differential
equations. Eq.\,(\ref{characteristics curves}) may be written as:
\begin{equation}\label{general solution curves}
 \frac{\mathrm{d} t}{1}= \frac{\mathrm{d} \Delta_t}{\gamma \Delta_t -
\frac{k}
 { M}}=-\frac{\mathrm{d} \rho(k,\Delta_t,t)}{\left(D_{pp}\Delta^2_t +
D_{xx}k^2 \right) \rho(k,\Delta_t,t)}\,.
\end{equation}
This pair of equations, valid on each characteristic curve,
enables a general solution of the partial differential equation
(\ref{delta k differential equation method of characteristics
form}) to be found as follows. We perform a first integration
using the first equality of Eq.\,(\ref{general solution curves}),
finding for one arbitrary constant $W$:
\begin{equation}\label{first arbitrary constant}
W=\mathrm{e}^{-\gamma t } \left({\Delta_t - \frac{k}{\gamma
M}}\right)\, ,
\end{equation}
from which it follows at $t=0$
\begin{equation}\label{first arbitrary constant 2}
W= {\Delta_0 - \frac{k}{\gamma M}}\, ,
\end{equation}
as $k$ is independent of time. Now we perform a second integration
using the second equality in Eq.\,(\ref{general solution curves}),
finding a second arbitrary constant $Z$:
\begin{eqnarray}\label{second arbitrary constant}
\fl Z=\frac{1}{\rho(k,\Delta_t,t)} \exp \left[ -\frac{D_{pp}
\Delta_t^2}{2 \gamma} - \frac{D_{pp}k \Delta_t}{M \gamma^2}
\right]
 \left({\Delta_t - \frac{k}{\gamma
M}}\right)^{-\frac{k^2}{\gamma}\big(D_{xx}+\frac{D_{pp}}{M^2
\gamma^2}\big)}\, ,
\end{eqnarray}
from which it follows at $t=0$
\begin{eqnarray}\label{second arbitrary constant 2}
\fl Z=\frac{1}{\rho(k,\Delta_0,0)} \exp \left[ -\frac{D_{pp}
\Delta_0^2}{2 \gamma} -\frac{D_{pp}k \Delta_0}{M \gamma^2} \right]
\left({\Delta_0 - \frac{k}{\gamma M}}\right)^{-\frac{k^2}{\gamma}
\big( D_{xx}+\frac{D_{pp}}{M^2 \gamma^2} \big)}\,.
\end{eqnarray}
By using Eq.\,(\ref{second arbitrary constant 2}) in
Eq.\,(\ref{second arbitrary constant}) we find for
$\rho(k,\Delta_t,t)$:
\begin{eqnarray}\label{ro kappa delta intermidiate}
\!\!\!\!\rho(k,\Delta_t,t)=&&\rho(k,\Delta_0,0)   \exp \left[
-\frac{D_{pp} \left(\Delta_t^2- \Delta_0^2\right)}{2 \gamma}
-\frac{D_{pp}k \left( \Delta_t - \Delta_0 \right)}{M \gamma^2}
\right]  \nonumber \\ \!\!\!\! && \times \left(\frac{{\Delta_t -
\frac{k}{\gamma M}}}{{\Delta_0 - \frac{k}{\gamma
M}}}\right)^{-\frac{k^2}{\gamma}\big(D_{xx}+\frac{D_{pp}}{M^2
\gamma^2}\big)}\,.
\end{eqnarray}
In order to express $\Delta_t$ as a function of $\Delta_0$ we use
Eqs.\,(\ref{first arbitrary constant}) and (\ref{first arbitrary
constant 2}), finding
\begin{equation}\label{Delta evolution}
\Delta_0= \Delta_t \mathrm{e}^{-\gamma t }+ \frac{k}{\gamma M}
\left(1 - \mathrm{e}^{-\gamma t }\right) \,.
\end{equation}
Substituting the previous expression for $\Delta_0$ in
Eq.\,(\ref{ro kappa delta intermidiate}), we finally obtain for
$\rho(k,\Delta_t,t)$:
\begin{eqnarray}\label{evoluzione matrice densità 0}
\rho(k,\Delta_t,t)=\rho\left(k,\Delta_t (1-\Gamma)+\frac{k}{\gamma
M}\Gamma,0\right)\times \\ \fl \exp  \left\{ \left[-
\left(D_{xx}+\frac{D_{pp}}{M^2 \gamma^2} \right)t
+\frac{D_{pp}}{M^2 \gamma^3} \left(\frac{\gamma^2}{2}+\Gamma
\right) \right]k^2 -\frac{D_{pp} \gamma^2}{M \gamma^2}k \Delta_t
-\frac{D_{pp}  \Gamma (2-\Gamma)}{2 \gamma}\Delta^2_t\,\right\}
  \, , \nonumber
\end{eqnarray}
where $\Gamma = 1-\exp (- \gamma t)$.

\section{}\label{Par:solving schrodinger cat dynamics}
In this appendix we obtain the time evolution of the
Schr\"{o}dinger cat initial state given in Eq.\,({\ref{Schrodinger
cat}}).

Computing the reduced density matrix $\rho(x,x',0)$ corresponding
to Eq.\,(\ref{Schrodinger cat}) and using Eq.\,(\ref{moving to
delta k}) to move to the $(k, \Delta_t)$ representation we obtain:
\begin{eqnarray}\label{schrodinger cat initial density matrix}
\fl \rho(k,\Delta_0,0)=\frac{1}{2\left(1+\exp
[-\frac{l^2}{8\sigma^2}]\right)} \left\{ \exp \left[
-\frac{\sigma^2}{2}k^2- \frac{\hbar^2}{8 \sigma^2}\Delta_0^2
+ i \frac{l}{2}k-i M v \Delta_0\right]+ \right. \nonumber \\
\fl\left. \exp \! \left[ - \frac{\sigma^2}{2}k^2- \frac{\hbar^2}{8
\sigma^2}\Delta_0^2 -i \frac{l}{2}k+ M v \Delta_0 \right]
 \! + \! \left(\exp \!
\left[- \frac{\sigma^2}{2}k^2- \frac{\hbar^2}{8
\sigma^2}\Delta_0^2- \frac{2 M v \sigma^2}{\hbar}k - \frac{\hbar
l}{4 \sigma^2} \Delta_0 \right] \right.\right. \nonumber \\ \fl
 \left.\left.+\exp  \left[-\frac{\sigma^2}{2}k^2-
\frac{\hbar^2}{8 \sigma^2}\Delta_0^2 +\frac{2 M v
\sigma^2}{\hbar}k + \frac{\hbar l}{4 \sigma^2} \Delta_0
\right]\right)  \times\exp \left[ - \frac{  2 M^2 v^2
\sigma^2}{\hbar^2}-\frac{l^2}{8 \sigma^2}\right] \right\}\,.
\end{eqnarray}
The form of this initial condition is of the kind:
\begin{equation}\label{schrodinger cat initial density matrix^2}
\rho(k,\Delta_0,0)=\sum_{j=1}^4
 \rho_j(k,\Delta_0,0)\, ,
\end{equation}
where comparing with Eq.\,(\ref{delta k density matrix}) we have
for the various coefficients the initial values
\begin{eqnarray}
\fl a^{1,2,3,4}_0=\frac{\sigma^2}{2},\quad b^{1,2,3,4}_0=0, \quad
c^{1,2,3,4}_0=\frac{\hbar^2}{8 \sigma^2}, \quad
d^1_0=-d^2_0=-\frac{l}{2}, \quad d^3_0=-d^4_0=-i \frac{2 M v
\sigma^2}{\hbar} , \nonumber  \\\fl e^1_0=-e^2_0=M v, \quad
e^3_0=-e^4_0=-i \frac{\hbar d}{4 \sigma^2} , \quad f^{1,2}_0=\ln
2\left[1-\exp \left(-\frac{d^2}{8\sigma^2}\right)\right] , \nonumber  \\
\fl f^{3,4}_0=\ln 2\left(1-\exp
\left[-\frac{d^2}{8\sigma^2}\right]\right) +\frac{2 M^2 v^2
\sigma^2}{\hbar^2}+\frac{l^2}{8 \sigma^2}\,.
\end{eqnarray}
By using Eq.\,(\ref{c coefficients evolution}) it is possible to
compute the time evolution of all the coefficients of the four
parts of Eq.\,(\ref{schrodinger cat initial density matrix^2}):
\begin{eqnarray}\label{evolution of coefficients}
\fl a^{1,2,3,4}_t=\bar{a}_t=\frac{\sigma^2}{2} +\frac{\hbar^2}{8
\sigma^2}\frac{\gamma^2}{M^2\gamma^2}
+\left(D_{xx}+\frac{D_{pp}}{M^2 \gamma^2} \right)t
-\frac{D_{pp}}{M^2 \gamma^3} \left(\frac{\Gamma^2}{2}+\Gamma
\right), \nonumber \\ \fl b^{1,2,3,4}_t=\bar{b}_t=
\frac{\hbar^2}{8 \sigma^2}\frac{ 2\Gamma (1-\Gamma)}{M\gamma}
+\frac{D_{pp} \Gamma^2}{M \gamma^2} , \quad
c^{1,2,3,4}_t=\bar{c}_t= \frac{\hbar^2}{8 \sigma^2} (1-\Gamma)^2+
\frac{D_{pp} \Gamma (2-\Gamma)}{2 \gamma} \nonumber ,
\\ \fl
 d^1_t=-d^2_t=\bar{d}_t=-\frac{l}{2}+   \frac{v\Gamma }{\gamma}, \quad
d^3_t=- d^4_t=\tilde{d}_t=-i\left(\frac{2 M v \sigma^2}{\hbar} +
\frac{\hbar l}{4 \sigma^2}\frac{\Gamma }{M
\gamma}\right),\nonumber
\\\fl e^1_t=-e^2_t=\bar{e}_t=M v (1-\Gamma), \quad
e^3_t=-e^4_t=\tilde{e}_t=-i \frac{\hbar l}{4 \sigma^2}
(1-\Gamma), \nonumber  \\
\fl f^{1,2}_t= f^{1,2}_0=\bar{f}_t ,\quad
f^{3,4}_t=f^{3,4}_0=\tilde{f}_t\,.
\end{eqnarray}
Then using Eq.\,(\ref{to coordinate representation}) we can move
to coordinate representation obtaining for the reduced density
matrix
\begin{eqnarray}\label{schrodinger cat final total density matrix}
\fl \rho(x,x',t)= \sum_{j=1}^4  \rho_{j}(x,x',t) =
\sum_{j=1}^4\exp  \!\left[-A^j_t(x-x')^2 - i B^j_t (x-x')(x+x') -
\right. \nonumber \\ \left.  C^j_t (x+x')^2 -i
D^j_t(x-x')-E^j_t(x+x')-F^j_t \right]
 ,
\end{eqnarray}
where
\begin{eqnarray}\label{schrodinger cate to coordinate representation}
\fl A^{1,2,3,4}_t=\bar{A}_t=\frac{4\bar{a}_t \bar{c}_t-
\bar{b}_t^2}{4 \hbar^2 \bar{a}_t}, \quad
B^{1,2,3,4}_t=\bar{B}_t=-\frac{ \bar{b}_t}{4 \hbar \bar{a}_t},
\quad
 C^{1,2,3,4}_t=\bar{C}_t=\frac{1}{16 \bar{a}_t}, \nonumber \\ \fl
D^1_t=
-D^2_t=\bar{D}_t=\frac{2\bar{a}_t\bar{e}_t-\bar{b}_t\bar{d}_t}{2
\hbar \bar{a}_t}, \quad  D^3_t \! = \! -D4_t \! =
\!\frac{2\bar{a}_t\tilde{e}_t-\bar{b}_t\tilde{d}_t}{2 \hbar
\bar{a}_t}\! =\! \tilde{D}_t, \,\,\,
 E^1_t=-E^2_t=\bar{E}_t= \frac{ \bar{d}_t}{4 \bar{a}_t}, \nonumber \\\fl
E^3_t=-E^4_t=\tilde{E}_t= \frac{ \tilde{d}_t}{4
 \bar{a}_t},\quad  \exp(-F^{1,2}_t)=\exp(-\bar{F}_t)=
\frac{\exp (-\bar{f}_t)}{2\sqrt{\pi \bar{a}_t}}\exp \!\!
\left[-\frac{\bar{d}_t^2}{4 \bar{a}_t}\right], \nonumber \\ \fl
\exp(-F^{3,4}_t)=\exp(-\tilde{F}_t)= \frac{\exp
(-\tilde{f}_t)}{2\sqrt{\pi \bar{a}_t}}\exp \!\!
\left[-\frac{\tilde{d}_t^2}{4 \bar{a}_t}\right].
\end{eqnarray}
Next from the last equation in Eq.\,(\ref{schrodinger cat final
total density matrix}) it follows that
\begin{eqnarray}\label{schrodinger cat final total density matrix^2}
\fl \rho(x,x',t)=  \exp  \!\!\left[-\bar{A}_t(x-x')^2 \!  - i
\bar{B}_t (x-x')(x+x')\!- \bar{C}_t (x+x')^2\right] \times
\nonumber
\\ \fl\left\{\exp \! \! \left[ -i
\bar{D}_t(x-x')-\bar{E}_t(x+x') -\bar{F}_t \right] \exp \! \!
\left[+i \bar{D}_t(x-x')+\bar{E}_t(x+x') -\bar{F}_t
\right]\nonumber \right.
\\ \fl +\left.
\exp \! \! \left[-i \tilde{D}_t(x-x')-\tilde{E}_t(x+x')
-\tilde{F}_t \right] + \exp  \!\!\left[+i
\tilde{D}_t(x-x')+\tilde{E}_t(x+x') -\tilde{F}_t \right] \right\}
 \,.
\end{eqnarray}
\\
Along the diagonal we have
\begin{eqnarray}\label{schrodinger cat final density matrix}
\fl\rho (x,x,t)=\frac{1}{2\left[1+\exp
(-\frac{l^2}{8\sigma^2})\right] 2 \sqrt{\pi \bar{a}_t}}   \left\{
\exp \left[  - \frac{(x+\bar{d}_t)^2}{4 \bar{a}_t}\right]+\exp
\left[ - \frac{(x-\bar{d}_t)^2}{4 \bar{a}_t} \right] \nonumber
\right.
\\ \left.
 \!\!\!\! + \exp  \! \left(- \frac{{x^2
-|\bar{d}_t|}^2}{4 \bar{a}_t} -\frac{2 M^2 v^2
\sigma^2}{\hbar^2}-\frac{l^2}{8 \sigma^2} \right) \cos
\frac{|\bar{d}_t| x}{2 \bar{a}_t} \right\}.
\end{eqnarray}

\end{document}